\documentclass[a4paper,11pt]{article}
\pdfoutput=1
\usepackage{jinstpub}

\title{A system for online beam emittance measurements and proton beam characterization}

\author[a,1]{K.~P.~Nesteruk, \note{Corresponding author.}}
\author[a]{M.~Auger,}
\author[a]{S.~Braccini,}
\author[a]{T.~S.~Carzaniga,}
\author[a]{A.~Ereditato,}
\author[a,b]{P.~Scampoli,}
\affiliation[a]{Albert Einstein Center for Fundamental Physics (AEC), \\ Laboratory for High Energy Physics (LHEP), University of Bern, \\Sidlerstrasse 5, CH-3012 Bern, Switzerland}
\affiliation[b]{Department of Physics ``E. Pancini'', University of Naples Federico II, \\ Complesso Universitario di Monte S. Angelo, I-80126, Naples, Italy}

\emailAdd{konrad.nesteruk@lhep.unibe.ch}

\abstract{A system for online measurement of the transverse beam emittance was developed. It is named $^{4}$PrOB$\varepsilon$aM (4-Profiler Online Beam Emittance Measurement) and was conceived to measure the emittance in a fast and efficient way using the multiple beam profiler method. The core of the system is constituted by four consecutive UniBEaM profilers, which are based on silica fibers passing across the beam. The $^{4}$PrOB$\varepsilon$aM system was deployed for characterization studies of the 18~MeV proton beam produced by the IBA Cyclone 18 MeV cyclotron at Bern University Hospital (Inselspital). The machine serves daily radioisotope production and multi-disciplinary research, which is carried out with a specifically conceived Beam Transport Line (BTL). The transverse RMS beam emittance of the cyclotron was measured as a function of several machine parameters, such as the magnetic field, RF peak voltage, and azimuthal angle of the stripper. The beam emittance was also measured using the method based on the quadrupole strength variation. The results obtained with both techniques were compared and a good agreement was found. In order to characterize the longitudinal dynamics, the proton energy distribution was measured. For this purpose, a method was developed based on aluminum absorbers of different thicknesses, a UniBEaM detector, and a Faraday cup. The results were an input for a simulation of the BTL developed in the MAD-X software. This tool allows machine parameters to be tuned online and the beam characteristics to be optimized for specific applications.}


\arxivnumber{1705.07486}

\begin{document}
\maketitle
\flushbottom

\section{\label{intro}Introduction}
Cyclotrons are nowadays common tools in medical applications and are used for the production of radioisotopes, as well as for cancer proton therapy. Medical cyclotrons have  a high scientific potential, well beyond the aim for which they were designed~\cite{cern_courier}. To exploit that, knowledge of the beam characteristics as a function of the main operational parameters is essential.
 
\begin{figure}[h]
\centering
\includegraphics[width=0.7\textwidth]{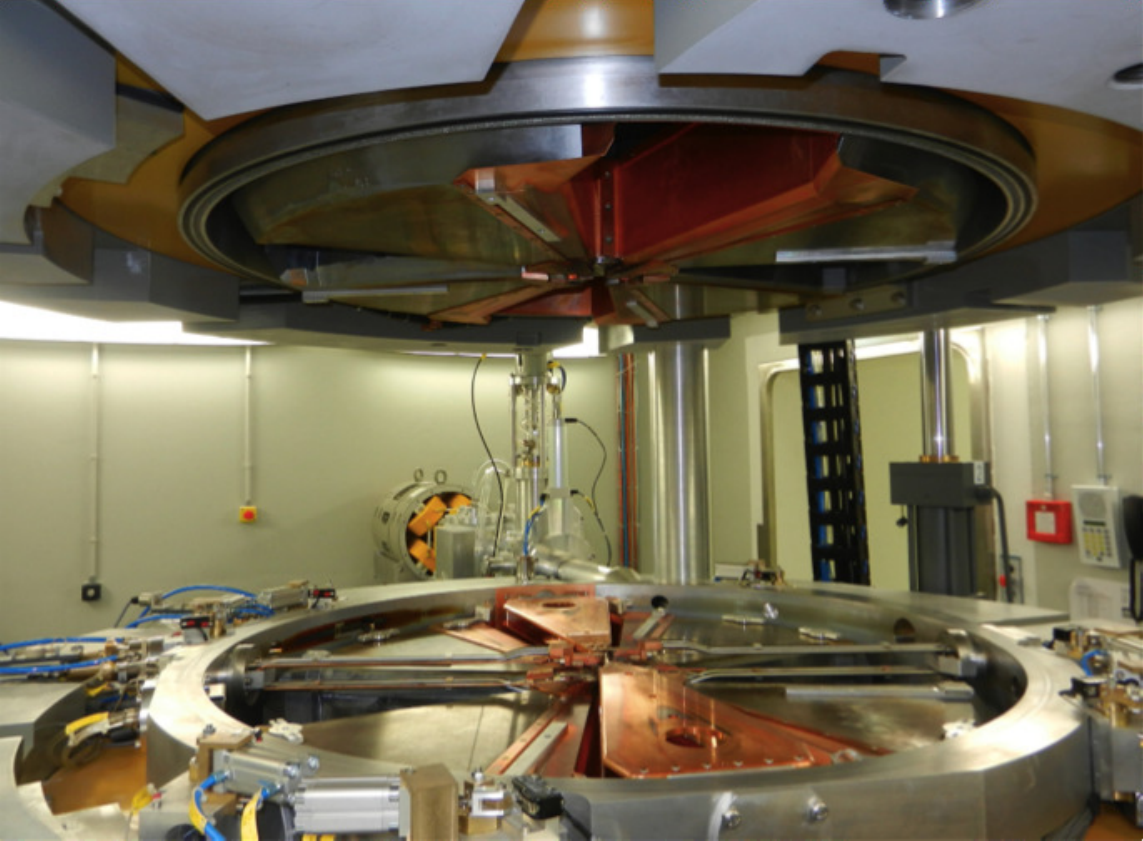}
\caption{\label{cyclotron} The Bern cyclotron opened during maintenance.}
\end{figure}

A cyclotron laboratory for radioisotope production and multi-disciplinary research is in operation at Bern University Hospital (Inselspital)~\cite{swan}. This facility hosts the IBA Cyclone 18~MeV proton cyclotron shown in figure~\ref{cyclotron}. This kind of machine is able to accelerate  H$^{-}$ and D$^{-}$ ions to the energy of 18~MeV and 9~MeV, respectively. However, to maximize the efficiency for daily medical radioisotope production in Bern, the machine is equipped with two H$^{-}$ ion sources. High beam currents up to 150~$\mu$A are provided in single or dual beam mode. Extraction is realized by stripping H$^{-}$ ions in a 5~$\mu$m thick pyrolytic carbon foil.

\begin{figure}[h]
\begin{center}
\includegraphics[width=0.7\textwidth]{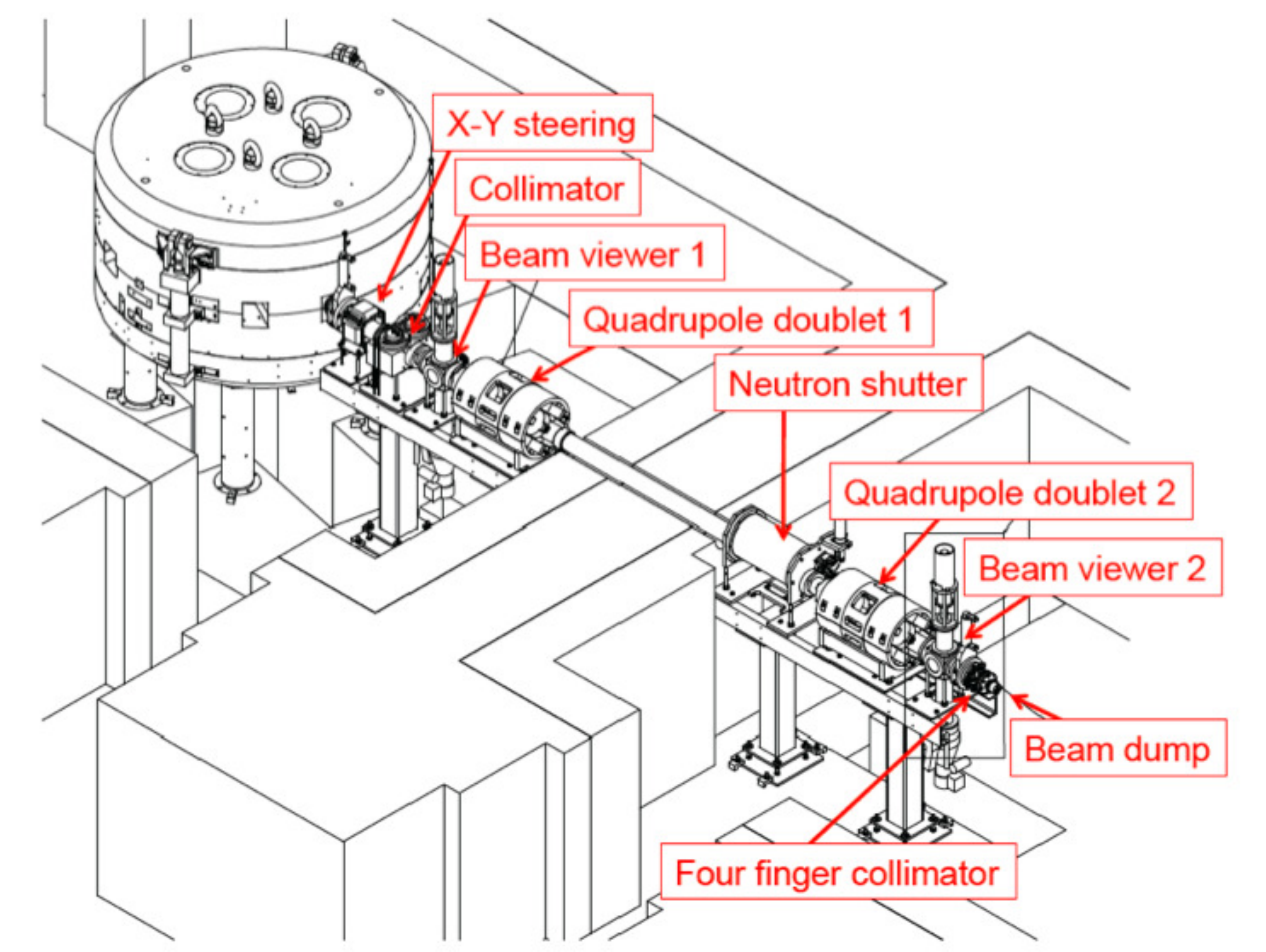}
\caption{\label{btl} Schematic view of the Bern cyclotron facility, where all the main components of the BTL are highlighted.}
\end{center}
\end{figure}

The Bern laboratory is equipped with a 6.5~m long Beam Transport Line (BTL), which is rare for a hospital based facility. It allows multi-disciplinary research to be carried out in parallel with daily radioisotope production. A schematic view of BTL is presented in figure~\ref{btl}. Alternate beam focusing and defocusing is realized by two horizontal-vertical quadrupole doublets. One is located in the cyclotron bunker, while the other in that of the BTL. A movable cylindrical neutron shutter is placed at the entrance of the BTL bunker to minimize the penetration of neutrons during routine production of radioisotopes. Experimental equipment used for scientific  activities, such as particle detectors or specific target stations, are installed at the end of the BTL. For several research activities performed with the BTL, low beam current intensities (down to pA range) are required. This is unusual for medical cyclotrons. However, the feasibility of stable operation of the Bern cyclotron at such low beam intensities was proven~\cite{low}. 

For beam profile measurements, beam monitors developed by our group and named UniBEaM are used~\cite{unibeam}. The UniBEaM detector is a compact device based on doped silica and optical fibers, which allows for fully automatized measurements of transverse beam profiles. A sensing fiber moves transversally across the beam and charged particles passing through the fiber cause scintillation. The produced light is transported to a read-out device, the signal is digitized and plotted online as a function of the fiber position. The precise measurement of the beam profile is at the basis of the study presented in this paper. The characterization of the cyclotron beam was performed by measuring both the transverse RMS beam emittance and the proton energy distribution. In particular, a system based on four UniBEaM detectors and named $^{4}$PrOB$\varepsilon$aM (4-Profiler Online Beam Emittance Measurement) was developed to provide an online measurement of the beam emittance. This system was employed for measurements of the transverse beam emittance as a function of the main cyclotron parameters. The experimental results were an input for a simulation of the BTL developed with the MAD-X software~\cite{madx}. This simulation is an important tool for optimizing irradiations for multi-disciplinary research. In this paper, we report on the first comprehensive study of the proton beam of the IBA 18 MeV cyclotron installed in Bern. It was performed for beam currents in the nA range, the typical operating conditions for research purposes. 

\section{\label{rms_emittance}Transverse RMS beam emittance}

The beam emittance is the main physical quantity used to characterize an accelerated particle beam~\cite{wiedemann}. In the case of transverse beam dynamics, the phase space is given by two variables for both horizontal and vertical planes: position ($x$ and $y$) and momentum ($p_{x}$ and $p_{y}$). The momenta are typically expressed by the angles $x'\approx p_{x}/p_{z}$ and $y'\approx p_{y}/p_{z}$, where $p_{z}$ is the longitudinal momentum. In the phase space ($x$, $x'$) or ($y$, $y'$), the points representing the particles are comprised inside an ellipse. The area of the phase space ellipse divided by $\pi$ is called the transverse beam emittance and is usually given in mm$\cdot$mrad. In further considerations, only the horizontal plane is discussed being the vertical plane completely equivalent. 

Realistic beams are usually far from being Gaussian and an appropriate statistical approach is required for a reliable estimation of the transverse beam emittance. In the case of an arbitrary density distribution $\rho(x, x')$, the following moments can be defined:
\begin{align}
\langle x^{2} \rangle &=\frac{\iint(x-\mu)^{2}\rho(x, x')dx'dx}{\iint\rho(x, x')dx'dx} \\
\langle x'^{2} \rangle &=\frac{\iint(x'-\mu ')^{2}\rho(x, x')dx'dx}{\iint\rho(x, x')dx'dx},     
\end{align}
and the covariance:
\begin{equation}
\langle xx' \rangle=\frac{\iint(x-\mu)(x'-\mu ')\rho(x, x')dx'dx}{\iint\rho(x, x')dx'dx},
\end{equation}
where $\mu$ and $\mu'$ are the expectation values for $x$ and $x'$, respectively. The beam matrix $\sigma(s)$ at the location $s$ along the beamline is therefore expressed in the following way:
\begin{align}
 \sigma(s) =
 \begin{pmatrix}
  \sigma_{11} & \sigma_{12} \\
  \sigma_{21} & \sigma_{22} \\
\end{pmatrix}
=
\begin{pmatrix}
  \langle x^{2}\rangle & \langle xx'\rangle \\
  \langle xx'\rangle & \langle x'^{2}\rangle \\
\end{pmatrix}.  
\end{align}
The RMS beam emittance $\varepsilon_{rms}$ is then given by the determinant of the $\sigma(s)$ matrix
\begin{equation}
\varepsilon_{rms}=\sqrt{\det(\sigma(s))},
\end{equation} 
and is independent of the location $s$, according to Liouville's theorem.

\section{Materials and methods}

\subsection{$^{4}$PrOB\texorpdfstring{$\boldmath{\varepsilon}$}{e}aM: a system for online measurement of the transverse beam emittance}

The $^{4}$PrOB$\varepsilon$aM system was conceived to measure the transverse beam emittance in a fast and efficient way using the multiple beam profiler method~\cite{mcdonald}. Unlike another common method based on quadrupole variation~\cite{mcdonald}, the use of multiple beam profilers does not require any prior knowledge of the optical elements of the beam transport. This technique is also simpler and more reliable than the ``pepper-pot'' method~\cite{mcdonald}. An advantage of the latter is that the shape of the beam in the horizontal and vertical phase spaces can be determined explicitly and an emittance plot showing contours of constant beam intensity can be generated for each. However, it is difficult to find the optimal conditions to obtain a decent precision of the measurement performed with the ``pepper-pot'' technique due to conflicting design considerations. For example, the spatial distribution is best determined by sampling the beam at small intervals (minimum hole spacing) but the angular distribution is more precisely determined as the spatial profiles of the non-overlapping beamlets get larger, maximizing the spacing.  

\begin{figure}[h!]
\centering
\includegraphics[width=0.7\textwidth]{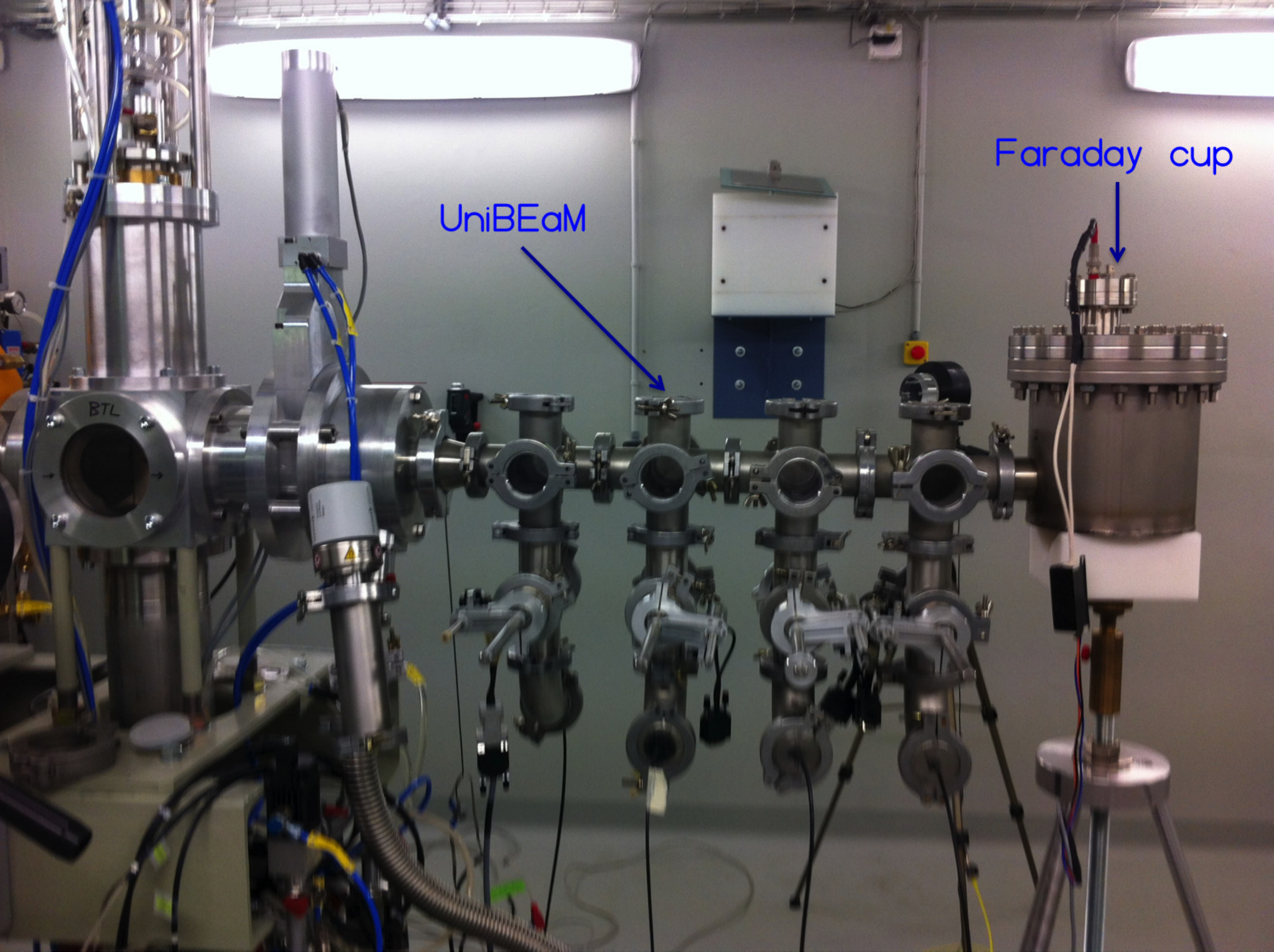}
\caption{The $^{4}$PrOB$\varepsilon$aM system and a Faraday cup installed on the BTL of the Bern cyclotron.}
\label{fig:4probeam}
\end{figure}

\begin{figure}[!htb]
   \centering
   \includegraphics[width=0.7\textwidth]{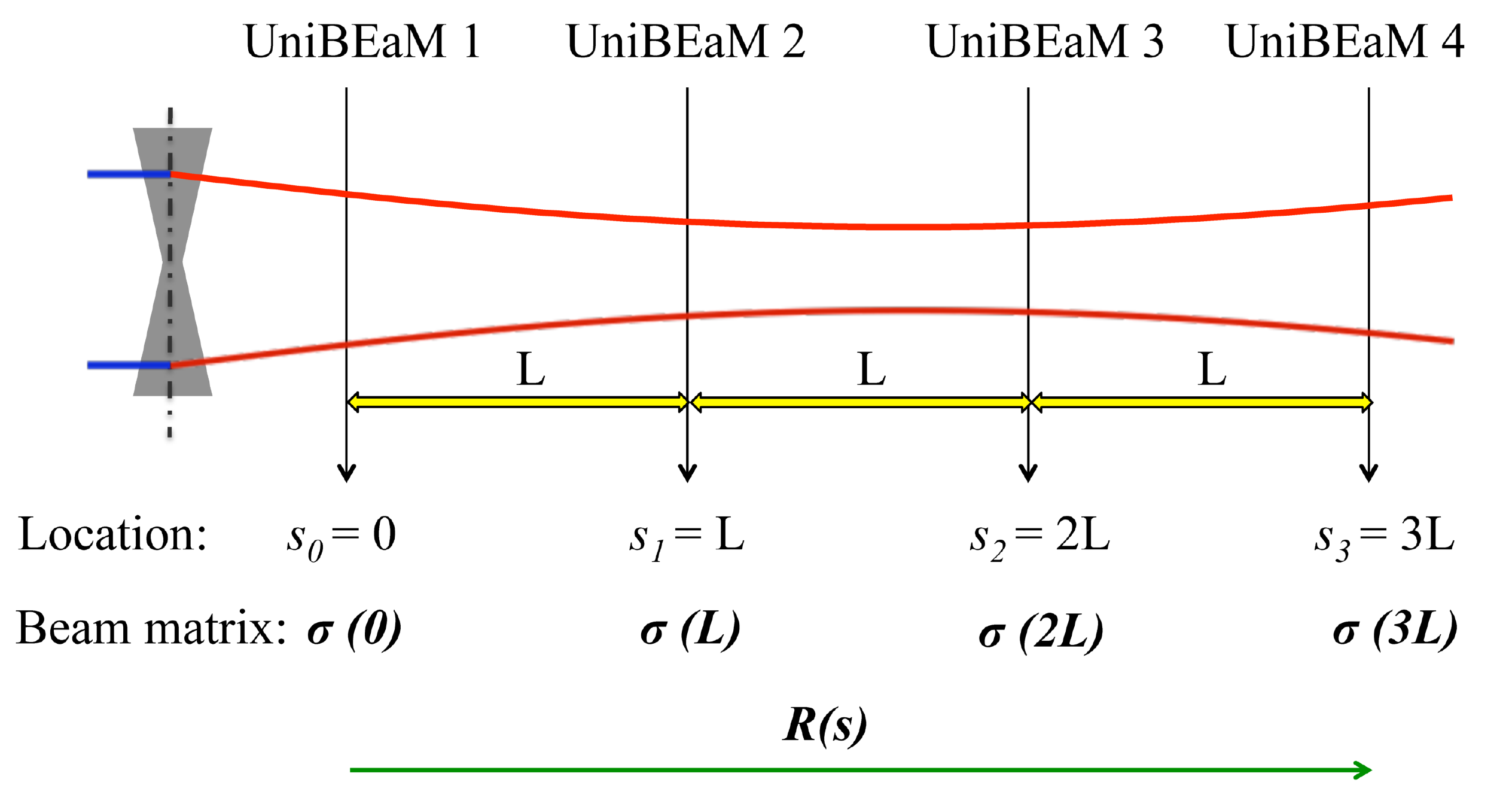}
   \caption{Sketch of principle of the multiple beam profiler method for the measurement of the transverse beam emittance.}
   \label{prof_method}
\end{figure}

\begin{figure}[!htb]
   \centering
   \includegraphics[width=\textwidth]{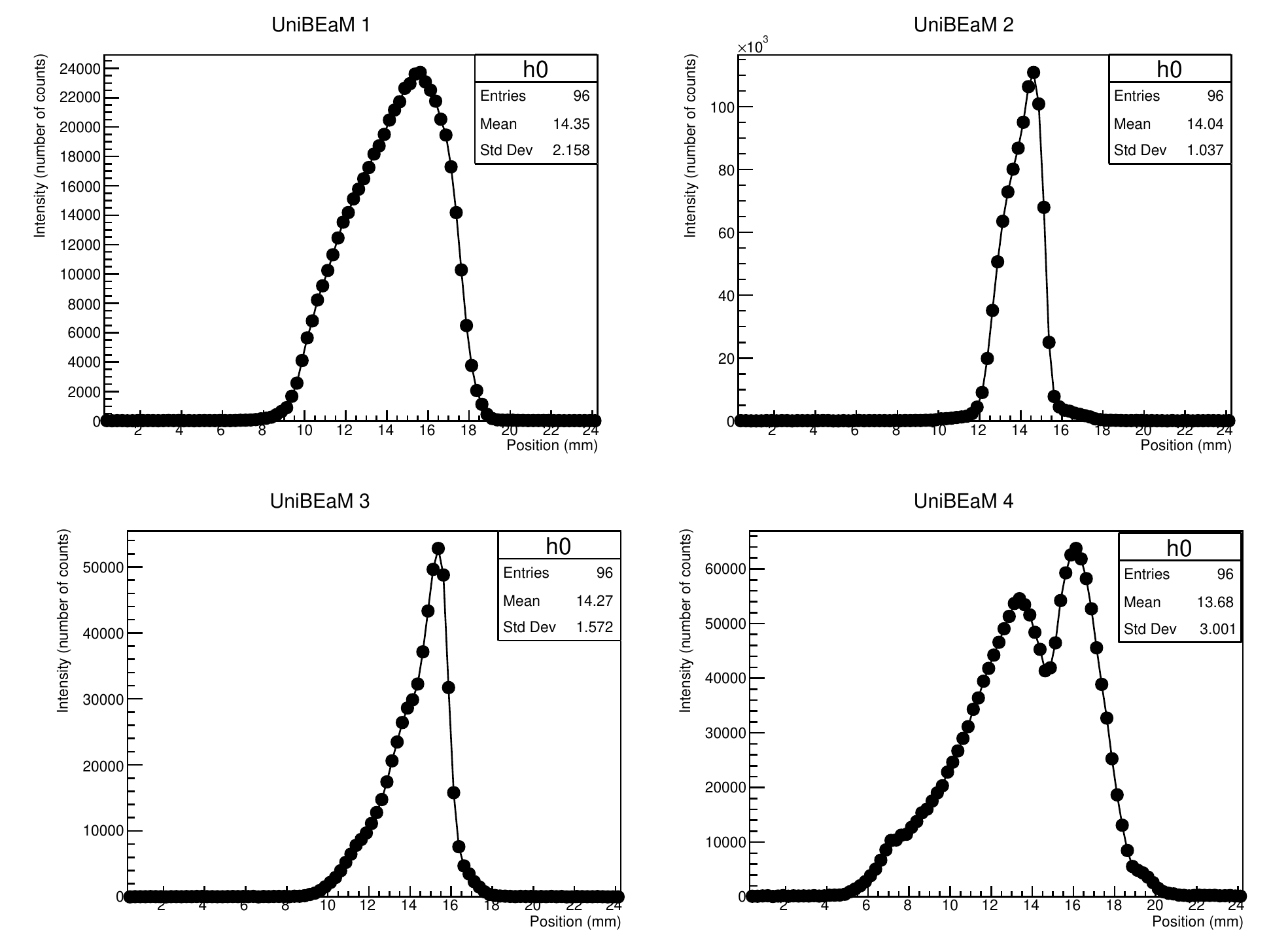}
   \caption{Beam profiles obtained for one measurement of the transverse beam emittance. The uncertainties are small, and therefore the error bars are not visible.}
   \label{profiles}
\end{figure}

The $^{4}$PrOB$\varepsilon$aM system consists of four UniBEaM detectors and its total length is 54~cm, which allows $^{4}$PrOB$\varepsilon$aM to be installed at nearly any location along beamlines or directly at the accelerator outport. For the measurements reported in this paper, the system was followed by a Faraday cup, which terminated the beamline to measure the beam current, as shown in figure~\ref{fig:4probeam}. For the measurements in the complementary plane, $^{4}$PrOB$\varepsilon$aM was rotated by 90 degrees.
All four beam profiles are measured simultaneously in order to minimize the influence of possible beam instabilities. Depending on the beam current range, different sensing fibers are used. For currents exceeding 1~nA, UniBEaMs are usually operated with non-doped optical fibers. Beam profiles are measured at four successive locations around a beam waist separated by a drift length $L=135$~mm, as depicted in figure~\ref{prof_method}. 
The optical signal from each detector is transmitted to a single-photon counter (commercial device of ID Quantique SA) and digitized. The whole data acquisition process consists of one full beam scan with a step of 0.25~mm over the maximum movement range of 24.25~mm and is controlled by a Raspberry~Pi~2 module with dedicated software. When the beam scan is complete, a ROOT~\cite{root} script is automatically launched, in which the whole data analysis is implemented. All four beam profiles are plotted in a separate window as histograms, as shown in figure~\ref{profiles}, and saved in a pdf file. 
The content of each histogram bin corresponds to the number of counts obtained over a period of 100~ms. The noise of the readout devices is subtracted using a measurement performed without beam in stable conditions. A single measurement of the beam emittance takes less than 1~minute. The variance and its uncertainty are calculated for each histogram of the profile giving an estimate of $\sigma_{11}(s)$  component of the beam matrix $\sigma(s)$ at the location $s$. The uncertainty of the estimated variance strongly depends on the beam profile integral. The uncertainties are smaller for higher light yields in the fiber corresponding to a larger total number of counts. Therefore, the smallest uncertainties are obtained for large beam currents and efficient scintillating fibers unless the read-out device becomes saturated. For emittance studies in a wide range of beam current, the non-doped optical fibers are chosen, which always operate below the saturation threshold and provide a good precision for currents exceeding 1~nA. The relative uncertainties can also vary between the four profiles, since the fibers used for the measurement are never equally efficient. 

The beam transfer matrix $R(s)$ involves only a drift:
\begin{equation}
R(s) =
\begin{pmatrix}
  1 & s \\
  0 & 1 
\end{pmatrix}.
\end{equation}
The beam matrix at any location $s$ with respect to the location of the first UniBEaM detector ($s_{0}=0$) is therefore given by the formula:
\begin{equation}
\sigma(s)=R(s)\sigma(0) R(s)^{T}.
\label{drift}
\end{equation}
From equation~(\ref{drift}) it can be derived that $\sigma_{11}(s)$ is a quadratic function of $s$:
\begin{align}
\sigma_{11}(s)&=\sigma_{22}(0)s^{2}+2s\sigma_{12}(0)+\sigma_{11}(0)\\
&=f(s;\sigma_{11}(0), \sigma_{12}(0), \sigma_{22}(0)), \nonumber
\end{align}
where $\sigma_{11}(0)$, $\sigma_{12}(0)$, and $\sigma_{22}(0)$ are the components of the $\sigma(0)$ matrix. These components and consequently the transverse emittance are evaluated by fitting the $f(s;\sigma_{11}(0), \sigma_{12}(0), \sigma_{22}(0))$ function using the four data points representing the estimated variance values as a function of the location $s$. The plot with the corresponding best fit is displayed and saved in a file. The precision of the method is the highest if the beam comes to a waist (a point of minimum spatial extent, where $\sigma_{12} = 0$), somewhere between the first and the last detector. The best precision is obtained when one of the detectors is located as close as possible with respect to a waist. The uncertainty of a single emittance measurement is calculated on the basis of the uncertainties of the fit parameters and of the correlations between them.
 
\subsection{\label{method_energy}Experimental method for measurement of the proton energy distribution}
\begin{figure}[!htb]
   \centering
   \includegraphics[width=0.65\textwidth]{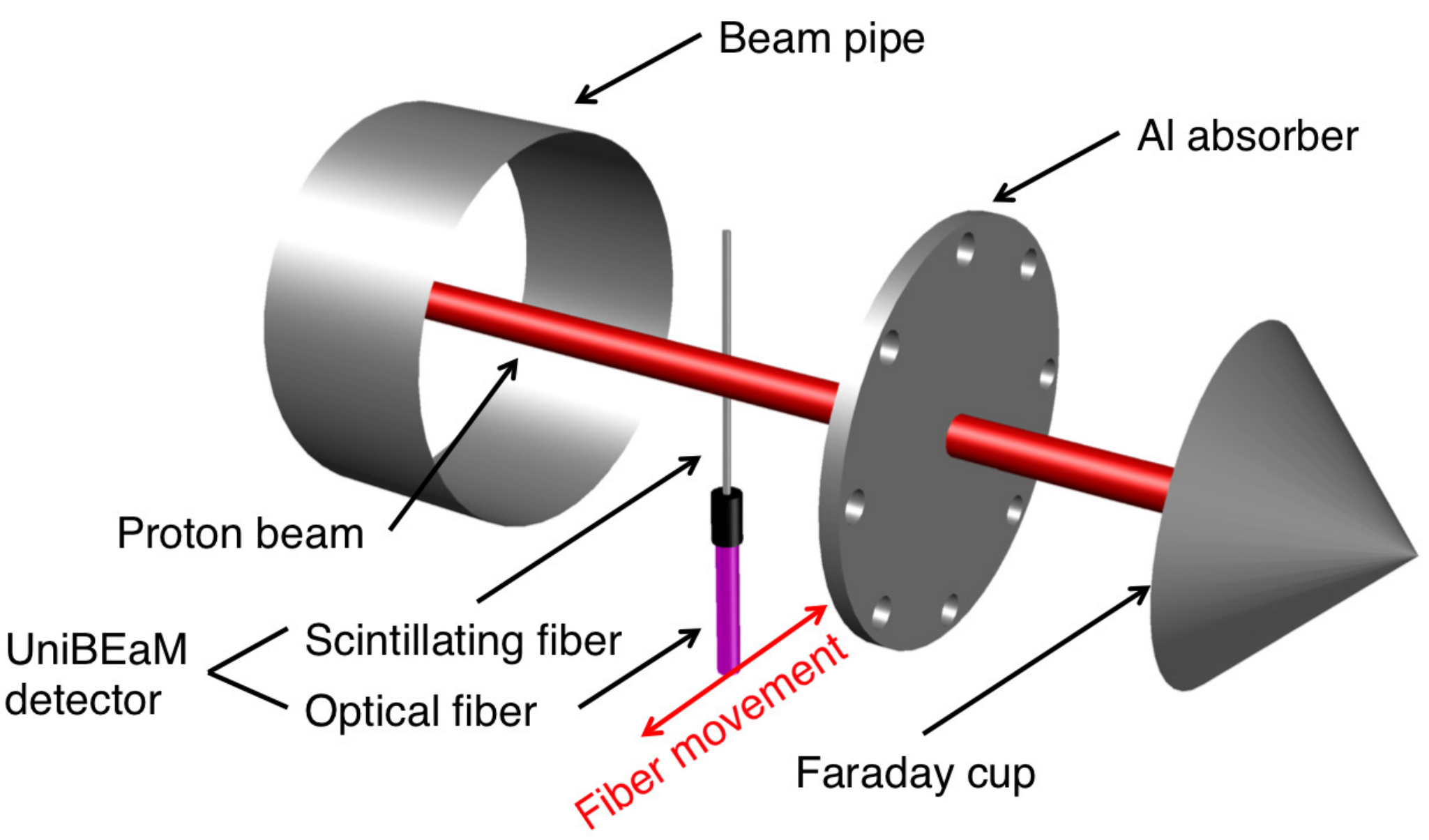}
   \caption{Schematic of the experimental method for the measurement of the proton energy distribution.}
   \label{energy_scheme}
\end{figure}

In order to characterize the energy distribution of the beam extracted to the BTL, a set of Al absorbers of different thickness was employed. Each thickness corresponded to a different maximum beam energy of the protons that could be stopped within the absorber, as reported in table~\ref{absorbers}. The maximum energy was determined by performing a SRIM~\cite{srim} simulation. For each absorber, a profile of the incident beam was measured by a UniBEaM detector and the intensity of the transmitted beam by a Faraday cup, as ilustrated in figure.~\ref{energy_scheme}. The UniBEaM detector was located in front of the absorber and was used to normalize the measured beam current. For the $i$-th absorber, the beam profile integral $S_{i}$ was calculated and the intensity $I_{i}$ of the transmitted beam was obtained from an electrometer. The profile integral holds a linear dependence on the beam current~\cite{unibeam}. A reference measurement was performed without any absorber, giving the integral $S_{0}$ and the beam current $I_{0}$. The fraction of the beam that was transmitted for each absorber $T_{i}$ is computed from the following formula:
\begin{equation}
T_{i}=\frac{S_{0}}{S_{i}}\cdot \frac{I_{i}}{I_{0}}\quad \quad i=1,2\dots,9.
\end{equation} 
The probability $P_{i,i+1}$ of finding a proton of the energy between the maximum proton energies corresponding to the absorbers $i$ and $i+1$ is given by the expression:
\begin{equation}
P_{i, i+1}=T_{i}-T_{i+1}\quad \quad i=1,2,\dots,8.
\end{equation}

\begin{table}[h]
\centering
\caption{\label{absorbers}%
Aluminum absorbers used for the measurement of the proton energy distribution.}
\begin{tabular}{ccc} \hline
\textbf{\#} &\textbf{Thickness [mm]} & \textbf{Max. proton energy [MeV]} \\ \hline
1 &  1.42 & 16.0 \\ 
2 & 1.50 & 16.5  \\
3 & 1.58 & 17.0 \\
4 & 1.67 & 17.5 \\
5 & 1.75 & 18.0 \\
6 & 1.84 & 18.5 \\
7 & 1.93 & 19.0 \\
8 & 2.02 & 19.5 \\
9 & 2.11 & 20.0 \\ \hline
\end{tabular}
\end{table}

\section{Characterization of the 18~MeV proton beam from the Bern medical cyclotron}
\subsection{Transverse RMS beam emittance as a function of cyclotron parameters}
For research purposes, the Bern cyclotron is operated in the manual mode, which allows the cyclotron parameters to be tuned in order to obtain beams according to specific needs.  These parameters include the main coil current, RF peak voltage, azimuthal stripper angle, and ion source arc current. For standard irradiations at the BTL, the values corresponding to the maximum beam transmission are used, which are reported in table~\ref{typical_settings}. For specific needs non-standard cyclotron parameters are chosen. In particular, the settings leading to non-optimal isochronism operation are deliberately selected when low current beams are required, as reported in~\cite{low}. To keep the same beam intensity over the course of long irradiations, the cyclotron parameters have to be continuously tuned. Therefore, the influence of a few important parameters on the beam emittance was studied by measuring the transverse RMS emittance as a function of the varied parameter. The $^{4}$PrOB$\varepsilon$aM system, able to make a single emittance measurement in a short time, was employed. Since in the majority of irradiations at the BTL the minimum value of the ion source arc current is used, this parameter was set to 1~mA for all the measurements reported. High values of the arc current are used for radioisotope production to maximize the beam intensity. However, for arc currents up to 100~mA, no significant change in the transverse beam emittance was observed. The value of the arc current influences the amount of plasma formed in the PIG (Penning Ionization Gauge) ion source.

\begin{table}[h]
\centering
\caption{\label{typical_settings}%
The standard settings of the cyclotron parameters used for research purposes.}
\begin{tabular}{cc} 
\hline
\textbf{Cyclotron parameter} & \textbf{Value} \\ \hline
Main coil & 136.9-137.2~A \\ 
RF peak voltage & 32 kV \\ 
Stripper angle & 84.2$^{\circ}$ \\ 
Ion source & 1~mA \\ \hline 
\end{tabular}
\end{table}

\noindent{\textbf{Variation of the main coil current}}

\begin{figure}[!htb]
   \centering
   \includegraphics[width=0.497\textwidth]{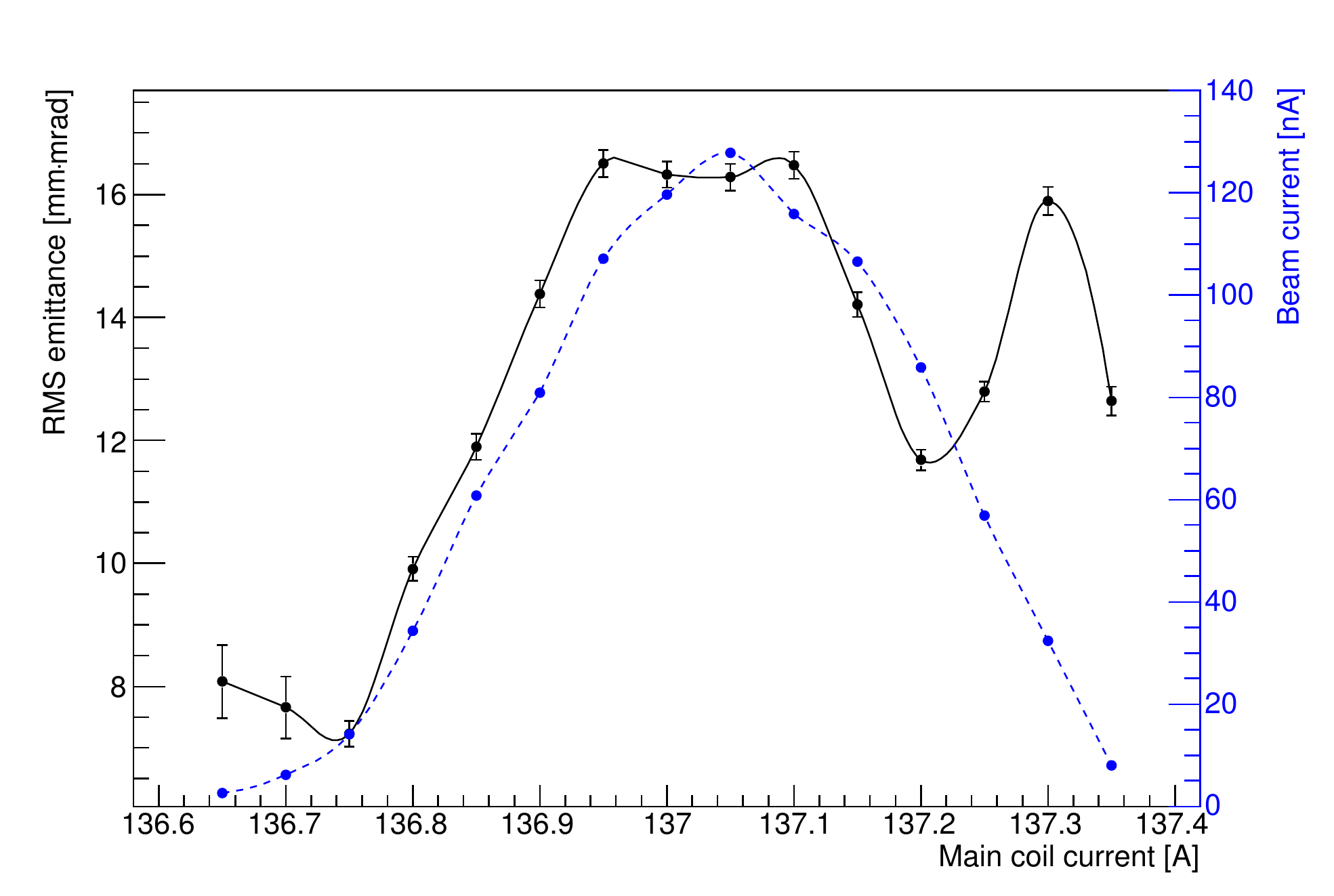}
     \includegraphics[width=0.497\textwidth]{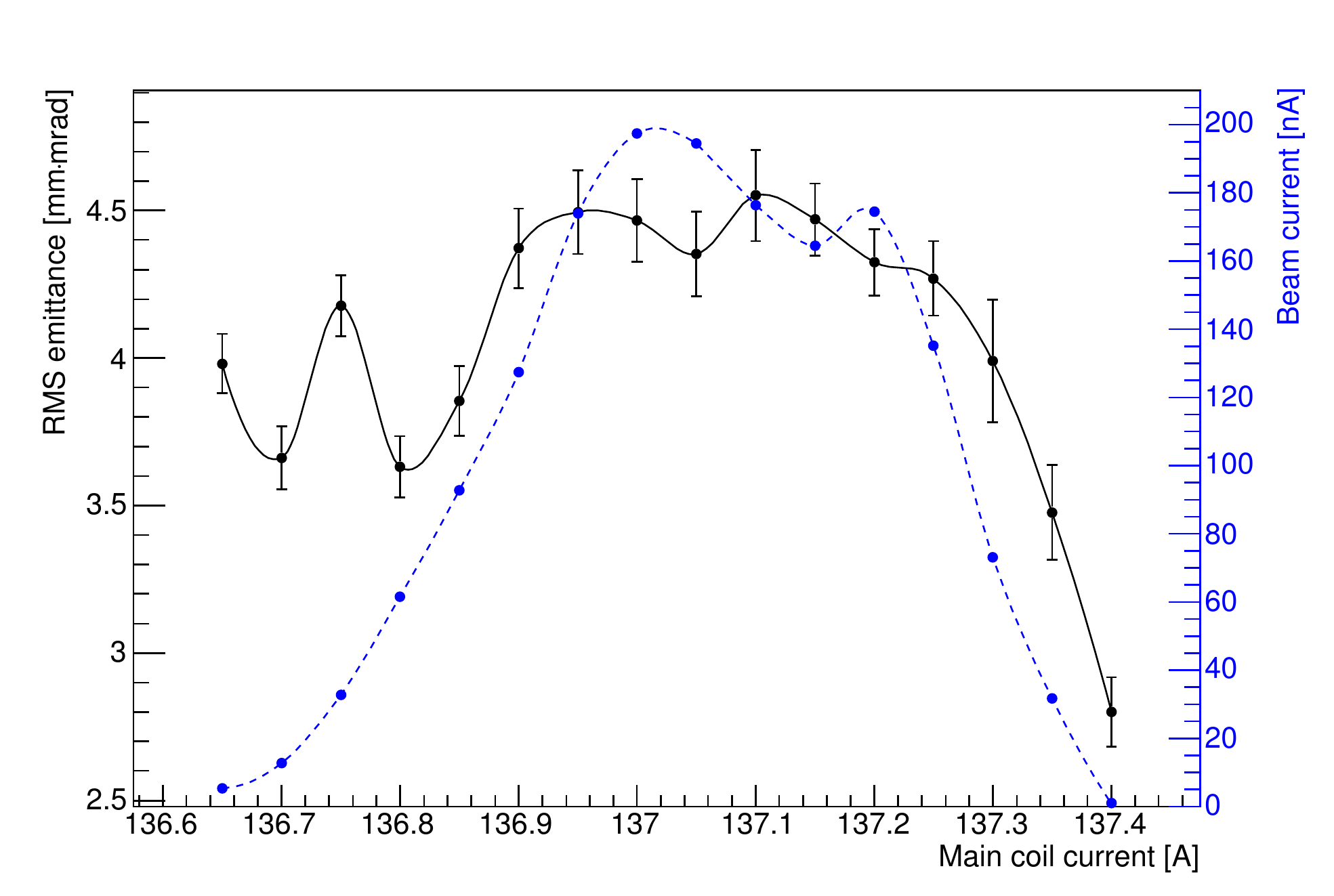}
   \caption{The horizontal (left) and vertical (right) transverse RMS emittance as a function of the main coil current for cold machine. The right vertical axis and blue dashed curve correspond to the beam current.}
      \label{emiMC}
\end{figure}

The current in the main coil is proportional to the magnetic field and thus influences isochronism. It is crucial for obtaining low intensity beams down to the pA range, since it allows the cyclotron to be operated in the regime of non-optimal isochronism. Moreover, the set point for the optimal isochronism has the tendency to drift towards higher values of the main coil current during long cyclotron runs. The influence of the main coil current on the transverse emittance was studied by gradually changing it and measuring the corresponding RMS emittance values. The other cyclotron parameters were set to the standard values (table~\ref{typical_settings}). The beam current was measured by means of a Faraday cup installed at the end of the beamline. The minimum and maximum main coil currents were determined by requiring a beam intensity of the order of 1~nA. The lowest value of the main coil current, which provides the beam intensity of about 1 nA, was chosen to be the lower edge value of the main coil current range. Further increase of the main coil current causes an increase of the beam intensity until the maximum has been reached. Afterwards, the beam intensity drops again and the highest value of the main coil current, which corresponds to the beam current of about 1 nA, was chosen to be the upper edge value of the main coil current range. In order to obtain a proper focusing, beam sizes, and location of a beam waist, the quadrupole settings used for the measurements in the horizontal plane were different than for the vertical one. Therefore, keeping exactly the same range of the beam current for both planes was not possible due to slightly different beam transmission for different quadrupole settings. Since the isochronism depends on the operation time, the main coil scans for horizontal and vertical planes were first performed for ``cold" machine, meaning that the cyclotron had not been operated before the measurement for at least a few hours. The results for the horizontal and vertical planes are presented in figure~\ref{emiMC}. In the case of the horizontal plane the emittance after an initial drop has a tendency to grow with the main coil current until the isochronism condition is reached. At the isochronism the emittance reaches a plateau corresponding to the region of the standard operation of the cyclotron. After that the emittance decreases reaching a local minimum. This is followed by another increase with a local maximum at a current of about 137.3~A. In this region the beam is unstable  and this effect may be due to the interception of the beam by the body of the second ion source or to a resonant condition producing beam losses, as reported in~\cite{low}. In the vertical plane, the variation of the emittance is much smaller and the plateau region is wider, as there is no acceleration in this plane. The measurements were performed several times in consecutive days giving reproducible results. The relative uncertainties in the vertical plane are larger than in the horizontal one mainly due to the different location of the beam waist and to the smaller variation of the beam size between the four profilers.

\begin{figure}[htb!]
   \centering
   \includegraphics[width=0.497\textwidth]{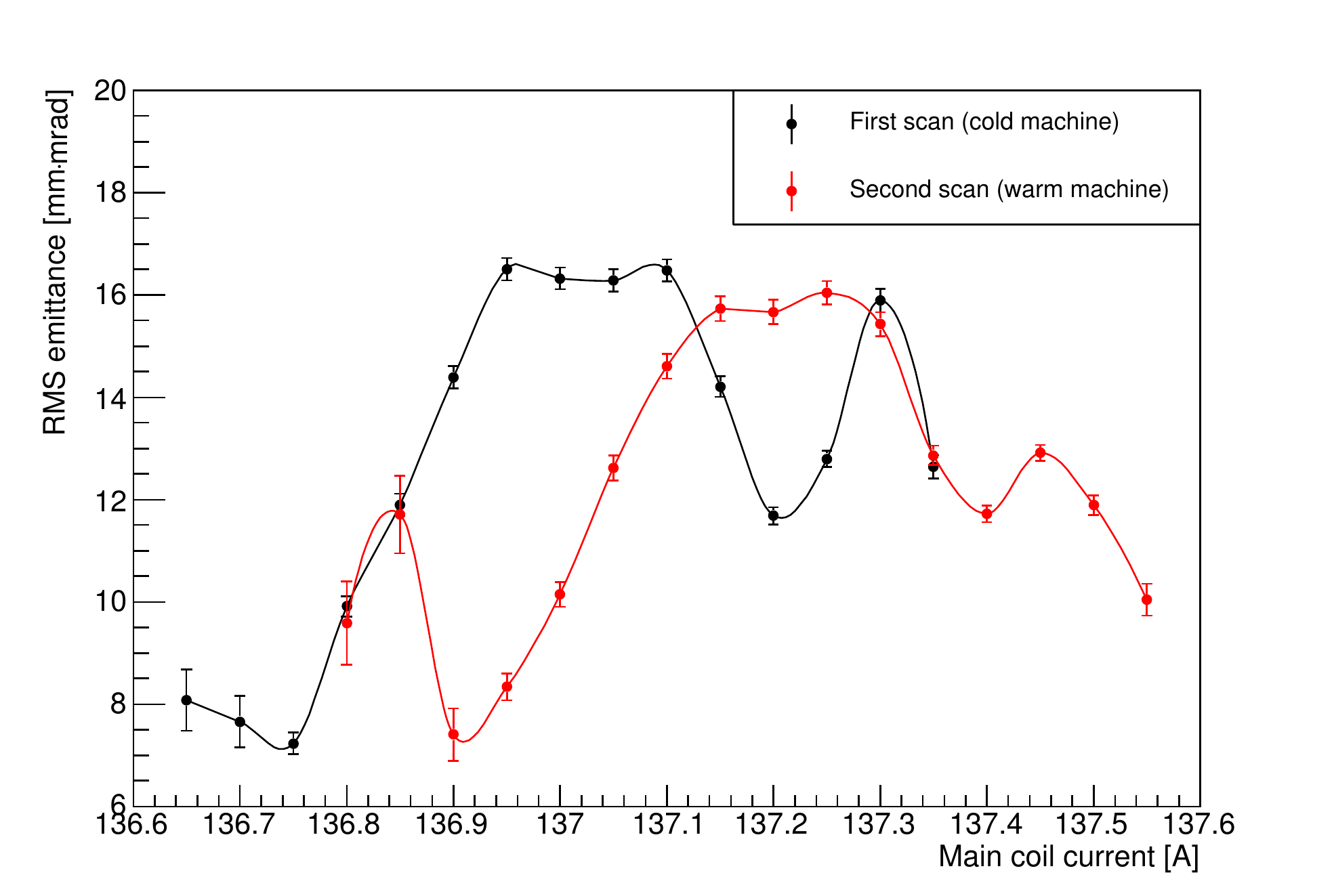}
      \includegraphics[width=0.497\textwidth]{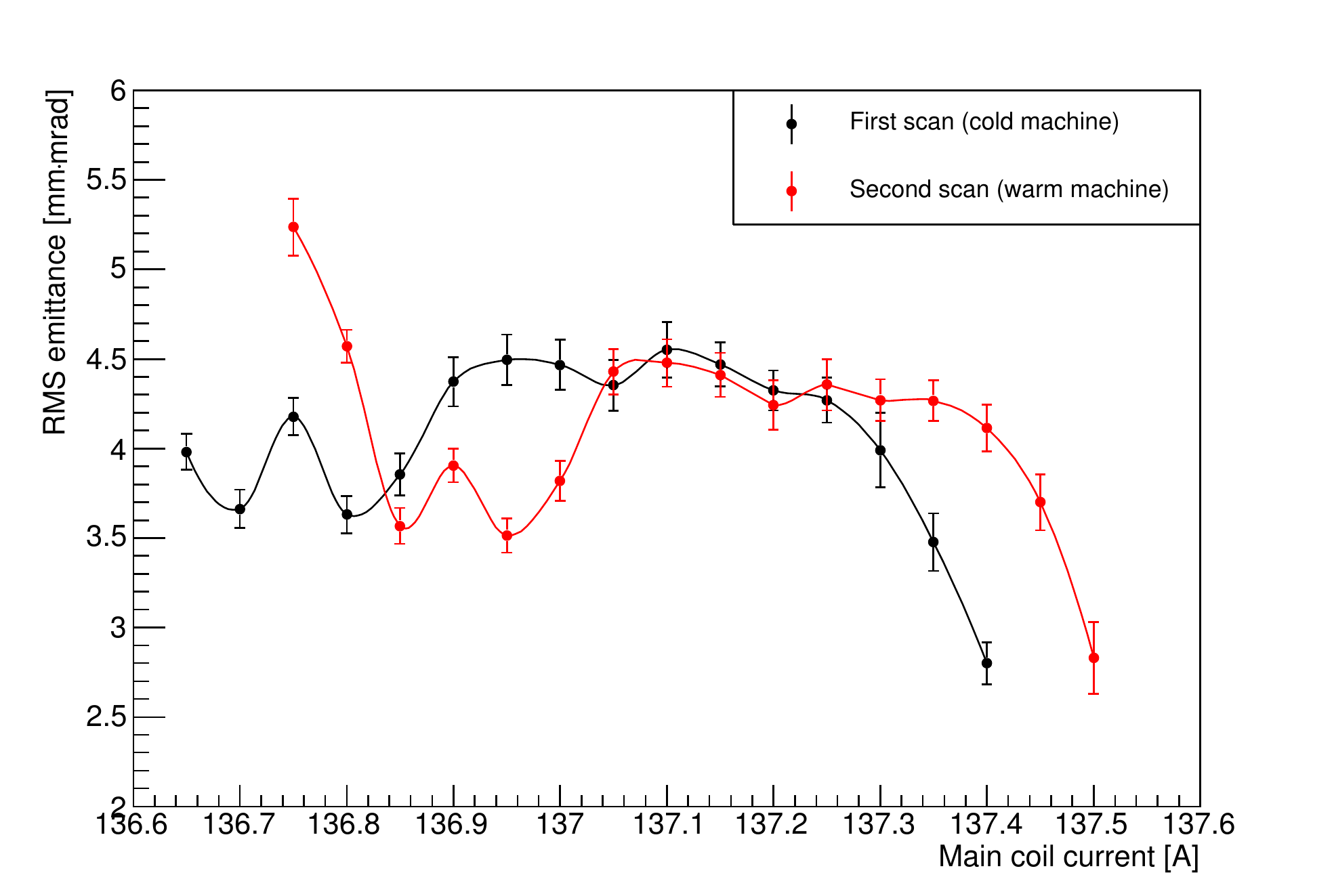}
   \caption{Comparison of the dependence of the horizontal (left) and vertical (right) RMS transverse emittance on the main coil current for cold and warm machine.}
   \label{comparison}
\end{figure}

The full main coil scan was repeated for both planes when the machine was ``warm", meaning that measurements were taken after a few hours of operation. The comparison of the results for the horizontal and vertical planes are presented in figure~\ref{comparison}. An offset of the curve corresponding to the second scan (warm machine) is clearly visible for both the horizontal and vertical planes. This offset is due to the warm-up of the machine showing the tendency of the optimal isochronism condition to drift towards higher values of the main coil current over the course of the cyclotron operation. The curve patterns for both scans present differences mostly for the extreme values of the main coil current. This is probably due to the fact that change in the range of the main coil operation extends measurable non-isochronous region. Furthermore, instabilities are observed for very low and very high values of the main coil current.

\noindent{\textbf{Variation of the RF peak voltage}}

\begin{figure}[!htb]
   \centering
   \includegraphics[width=0.497\textwidth]{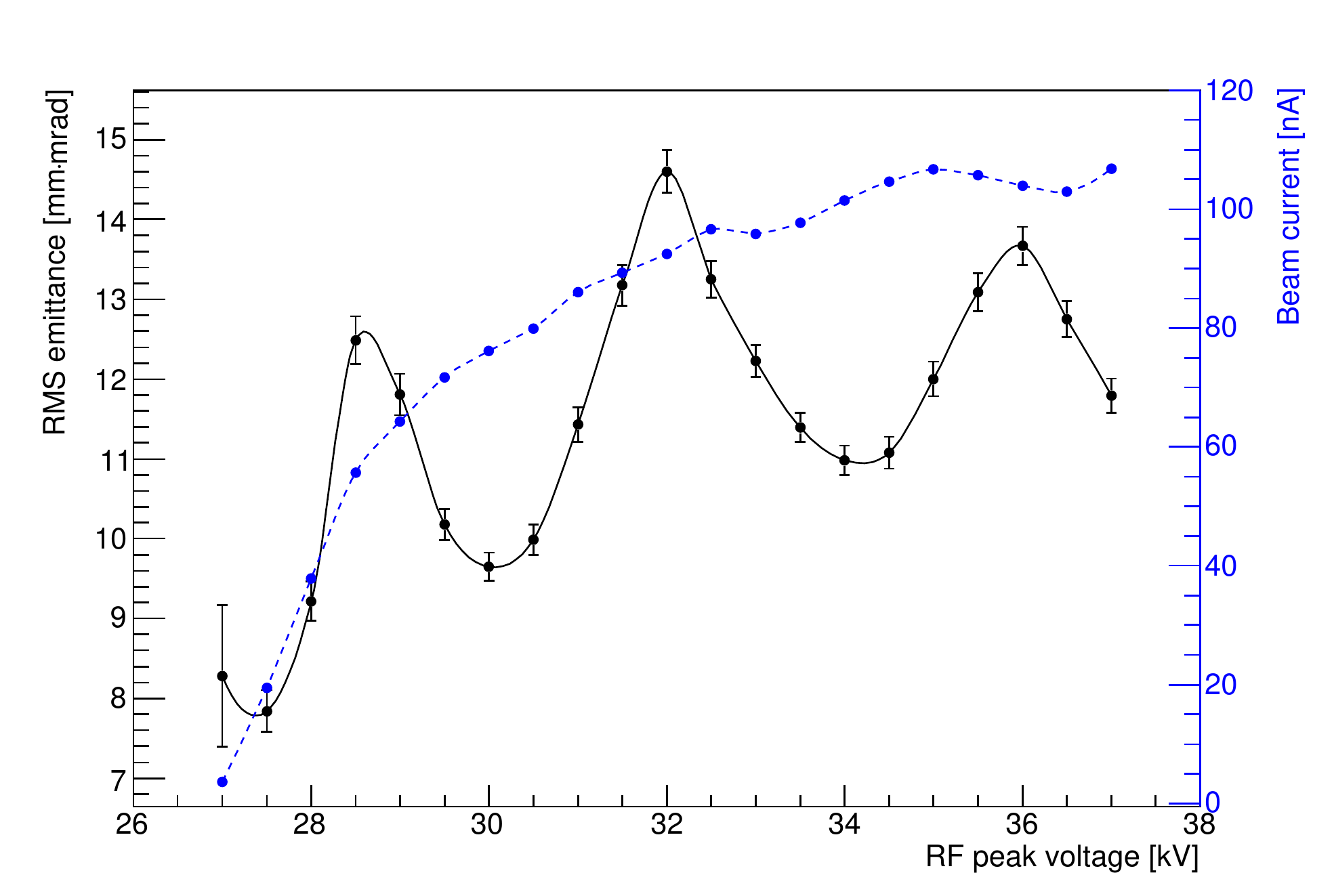}
      \includegraphics[width=0.497\textwidth]{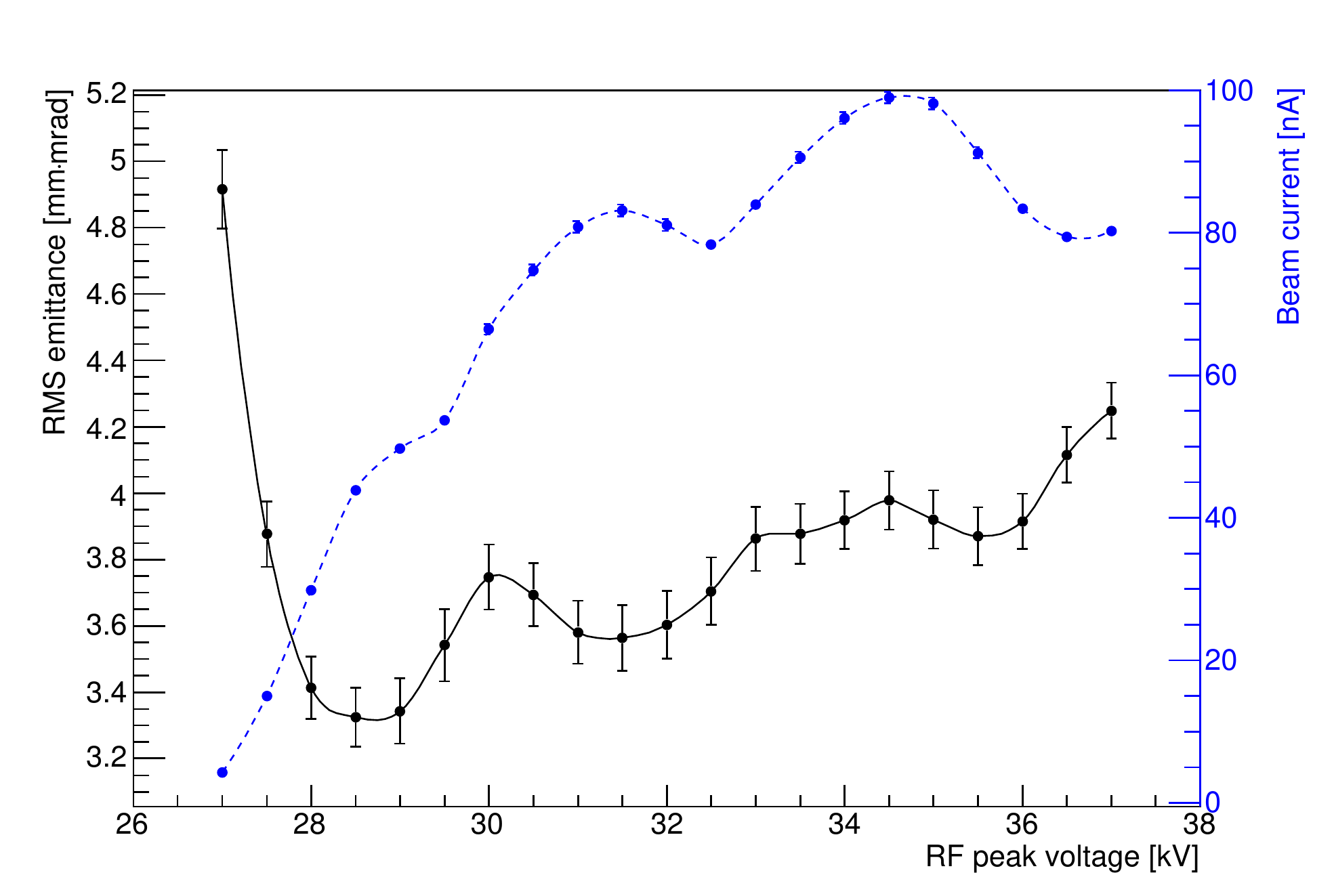}
   \caption{The horizontal (left) and vertical (right) transverse RMS emittance as a function of the RF peak voltage. The right vertical axis and blue dashed curve correspond to the beam current.}
   \label{emiRF}
\end{figure}

The peak voltage of the 42~MHz RF system is responsible for acceleration and extraction of H$^{-}$ ions from the chimney of the ion source, and  can be varied from 27~kV to 37~kV. The RF peak voltage was varied in the full range in steps of 0.5~kV. The beam current was again monitored by means of the Faraday cup. The main coil current was set to 137.05~A and 137.00~A for the horizontal and vertical planes, respectively. These values of the main coil current provided operation in the region of the emittance stability (figures~\ref{emiMC}~and~\ref{comparison}). The results for the horizontal and vertical planes are presented in figure~\ref{emiRF}. In both planes oscillations of the emittance occur, while the emittance values at the local minima tend to increase. An exact explanation of the observed effects is very difficult, since the centering of the beam is unknown, as there is no differential radial probe inside the machine. Also, the phase of the radial betatron oscillations is not controlled. The changes of the emittance measured at the chosen stripper azimuthal angle are caused by a superposition of the RF voltage modification, beam off-centering, and phase of the betatron oscillations. The total emittance of the extracted beam is a sum of emittances of beam parts extracted at the turns $N, N+1, N+2,...$. One may expect that higher RF voltages reduce the number of turns during acceleration and at the same time enlarge the accepted range of initial RF phases passing from the ion source to the stripper foil. This may be the reason for observing slightly larger beam emittances for larger values of the RF peak voltage.

\noindent{\textbf{Variation of the stripper angle}}

\begin{figure}[!htb]
   \centering
   \includegraphics[width=0.497\textwidth]{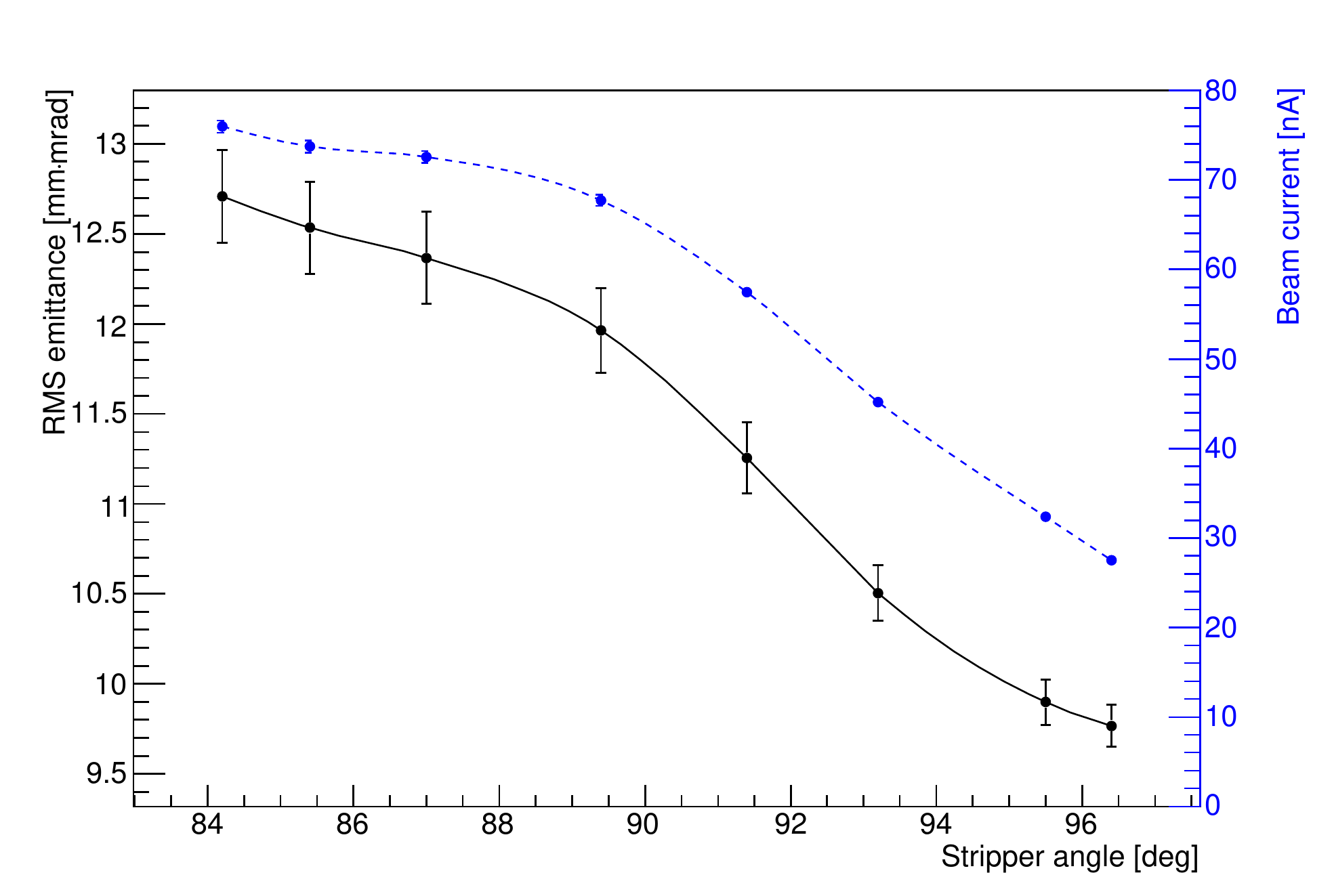}
      \includegraphics[width=0.497\textwidth]{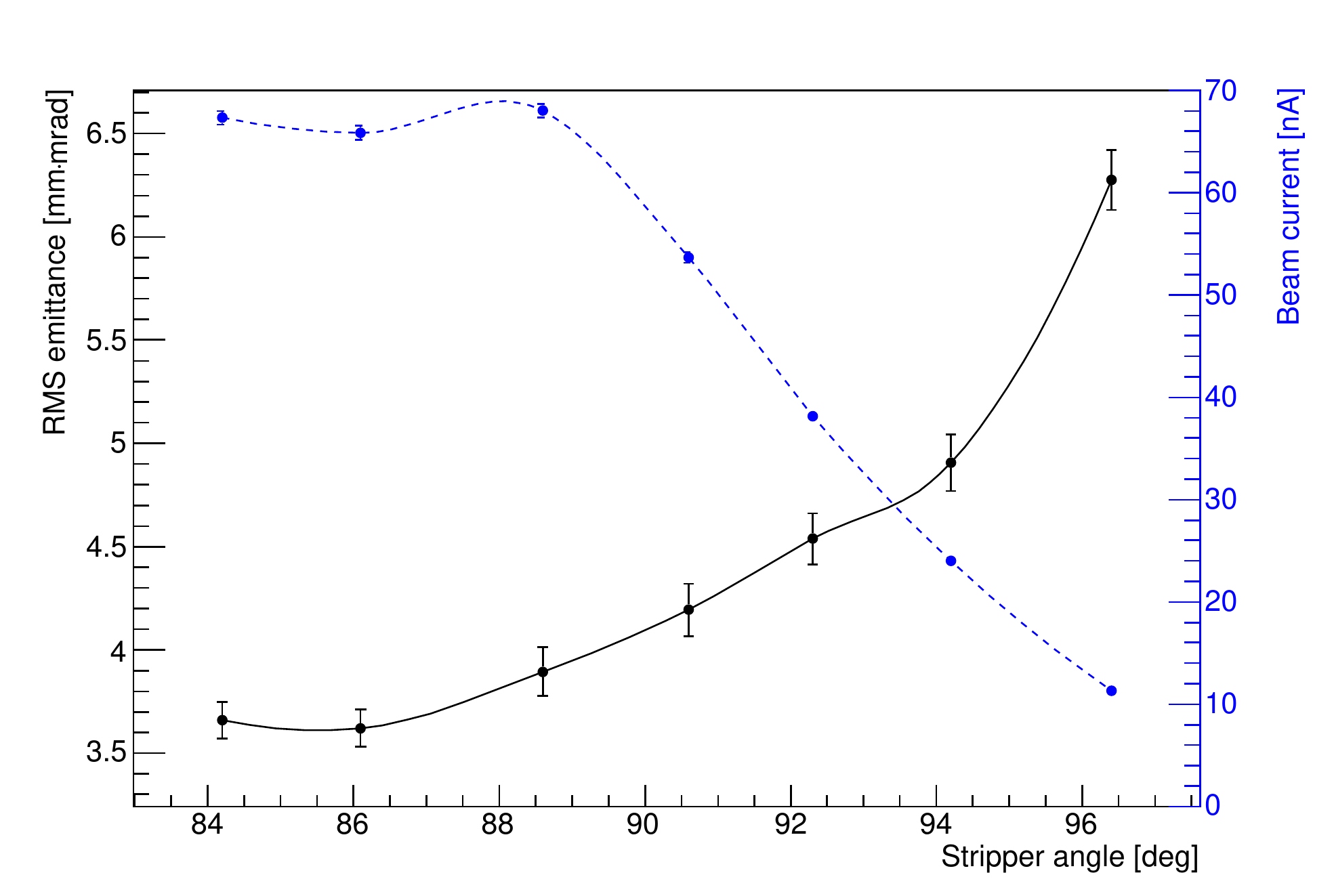}
   \caption{The horizontal (left) and vertical (right) transverse RMS emittance as a function of the stripper angle. The right vertical axis and blue dashed curve correspond to the beam current.}
   \label{emiStripper}
\end{figure}

The stripper angle can be adjusted to optimize the extracted proton beams. In most irradiations a default stripper angle of 84$^{\circ}$ is used. The optimal angle can change due to stripper deformation or when a new stripper is installed during periodical cyclotron maintenance. The azimuthal angle was varied from the nominal value to 96.4$^{\circ}$. The RF peak voltage and the main coil current were set for these measurements to 32~kV and 137.05~A, respectively. Since the nominal value of the stripper angle is chosen so that the beam intensity is maximum, the beam current monitored by the Faraday cup was decreasing with the increase of the angle. The transverse RMS emittance was found to decrease in the horizontal plane (figure~\ref{emiStripper}~(left)) and increase in the vertical one (figure~\ref{emiStripper}~(right)). The interpretation of the obtained result is difficult. It is likely that these changes of the beam emittance would not be observed if the beam was well centered and passed through the stripper in a single turn. Therefore, the results indicate imperfections in the studied machine.

\subsection{Transverse RMS beam emittance for the standard cyclotron settings}
The transverse RMS beam emittance was measured with two different methods for the cyclotron settings typically used for multi-disciplinary research with the BTL~(table~\ref{typical_settings}). The first technique employed was the variation of the quadrupole strength~\cite{mcdonald, mostacci}. For this method, the last quadrupole magnet of the BTL, which is defocusing in the horizontal plane and focusing in the vertical, was varied. The corresponding beam profiles at a distance of 694~mm from the quadrupole were measured with the UniBEaM detector for each magnet setting. The second technique was the use of multiple beam profilers, for which the $^{4}$PrOB$\varepsilon$aM system was again used. During the measurements of the transverse beam emittance, the cyclotron parameters were kept constant and set to the standard values. The beam current, as monitored by means of a Faraday cup, was about 250~nA.

\begin{figure}[!htb]
   \centering
   \includegraphics[width=0.7\textwidth]{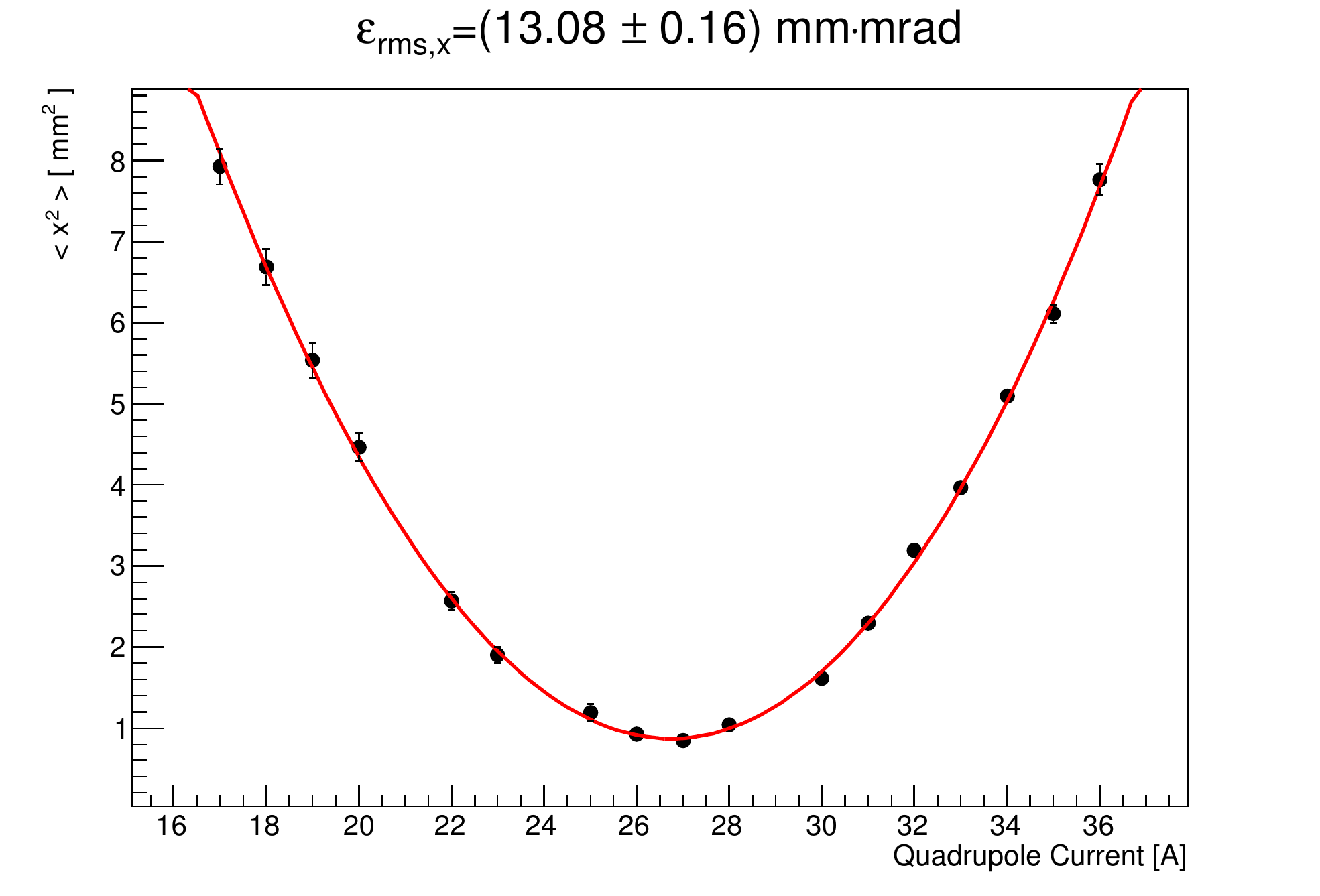}
   \caption{Variance as a function of the quadrupole current obtained in the horizontal plane. The red line corresponds to the best fit.}
   \label{var_x}
\end{figure}

The estimated variance in the horizontal plane $\langle x^{2} \rangle$ as a function of the quadrupole current together with the fitted curve for the quadrupole variation method is shown in figure~\ref{var_x}. A similar curve was obtained for the vertical plane, as presented in our preliminary study~\cite{IBIC}. Fit results for both planes and the corresponding emittance values are listed in table~\ref{var_results}. The transverse RMS emittance in the horizontal plane is $3.6$ times larger than the one in the vertical plane.

\begin{table}[hbt]
   \caption{Fit parameters and the RMS emittance values obtained by quadrupole variation for both horizontal and vertical planes.}
    \centering 
   \begin{tabular}{lcc} \hline
       \textbf{Fit parameter} & \textbf{Horizontal plane} & \textbf{Vertical plane} \\
       \hline
           $\sigma_{11}$ [mm$^{2}$]         & $200.23 \pm 0.08$            & $21.59 \pm 0.36$     \\
           $\sigma_{12}$ [mm$\cdot$mrad]    & $-322.66 \pm 0.08$            & $-2.98 \pm 0.07$        \\
           $\sigma_{22}$ [mrad$^{2}$]       & $520.80\pm0.22$             &   $1.02 \pm 0.02$ \\
       \hline
       $\tilde{\chi}^2$  & $0.98$ &  $1.04$ \\
       \hline
       \hline
       $\varepsilon_{rms}$ [mm$\cdot$mrad]  & $13.08\pm0.16$  &    $3.63 \pm 0.04$     \\  \hline
   \end{tabular} 
   \label{var_results}
\end{table}

\newpage

\begin{figure}[!htb]
   \centering
   \includegraphics[width=0.7\textwidth]{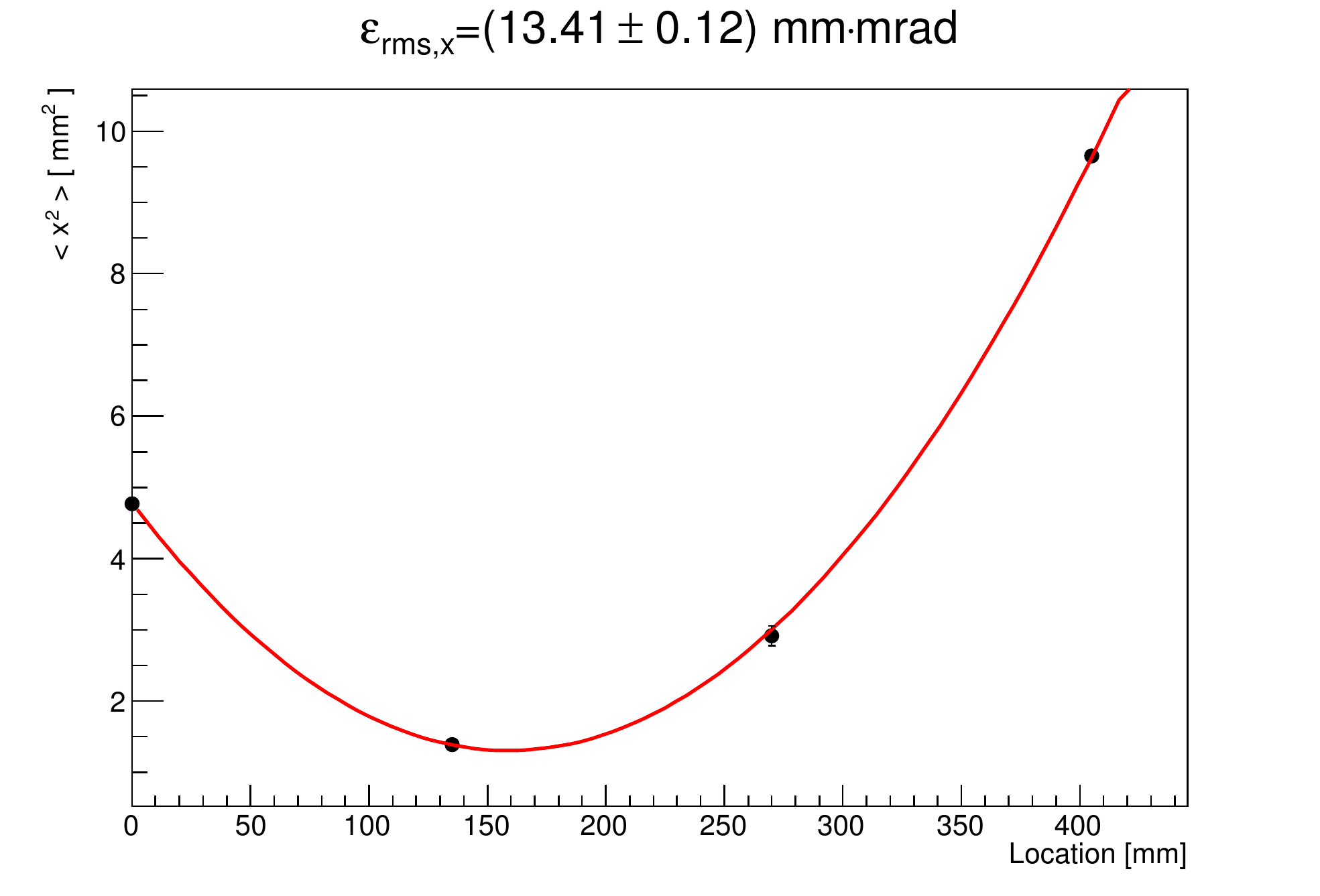}
   \caption{Variance as a function of the location obtained in the horizontal plane. The red line corresponds to the best fit.}
   \label{prof_x}
\end{figure}

The estimated variance in the horizontal plane $\langle x^{2} \rangle$ as a function of the location $s$ together with the fitted curve for the multiple beam profiler method is shown in figure~\ref{prof_x}. A similar curve was obtained for the vertical plane, as reported in~\cite{IBIC}. Fit results for both planes and the corresponding emittance values are given in table~\ref{prof_results}.

\begin{table}[hbt]
   \centering
   \caption{Fit parameters and the RMS emittance values obtained by using multiple profilers for both horizontal and vertical planes.}
   \begin{tabular}{lcc} \hline
       \textbf{Fit parameter} & \textbf{Horizontal plane} & \textbf{Vertical plane} \\
       \hline
           $\sigma_{11}$ [mm$^{2}$]         & $4.79 \pm 0.09$            & $0.75 \pm 0.04$     \\
           $\sigma_{12}$ [mm$\cdot$mrad]    & $-21.90 \pm 0.48$            & $-1.06 \pm 0.19$        \\
           $\sigma_{22}$ [mrad$^{2}$]       & $137.72\pm2.06$             &   $17.99 \pm 1.15$ \\
       \hline
        $\tilde{\chi}^2$  & $0.47$ &  $0.76$ \\
       \hline
       \hline
       $\varepsilon_{rms}$ [mm$\cdot$mrad]  & $13.41\pm0.12$  &    $3.53 \pm 0.13$     \\ \hline
   \end{tabular}
   \label{prof_results}
\end{table}

The results obtained by employing the two methods were found to be in agreement within 1.65$\sigma$ and 0.71$\sigma$ for the horizontal and vertical planes, respectively. The transverse RMS beam emittance in the horizontal plane is almost 4 times larger than the one in the vertical plane. This is typical for cyclotrons where acceleration takes place in the horizontal plane. This causes an increase of the particle position spread along the $x$-direction.

\subsection{Proton energy distribution}
\begin{figure}[!t]
   \centering
   \includegraphics[width=0.7\textwidth]{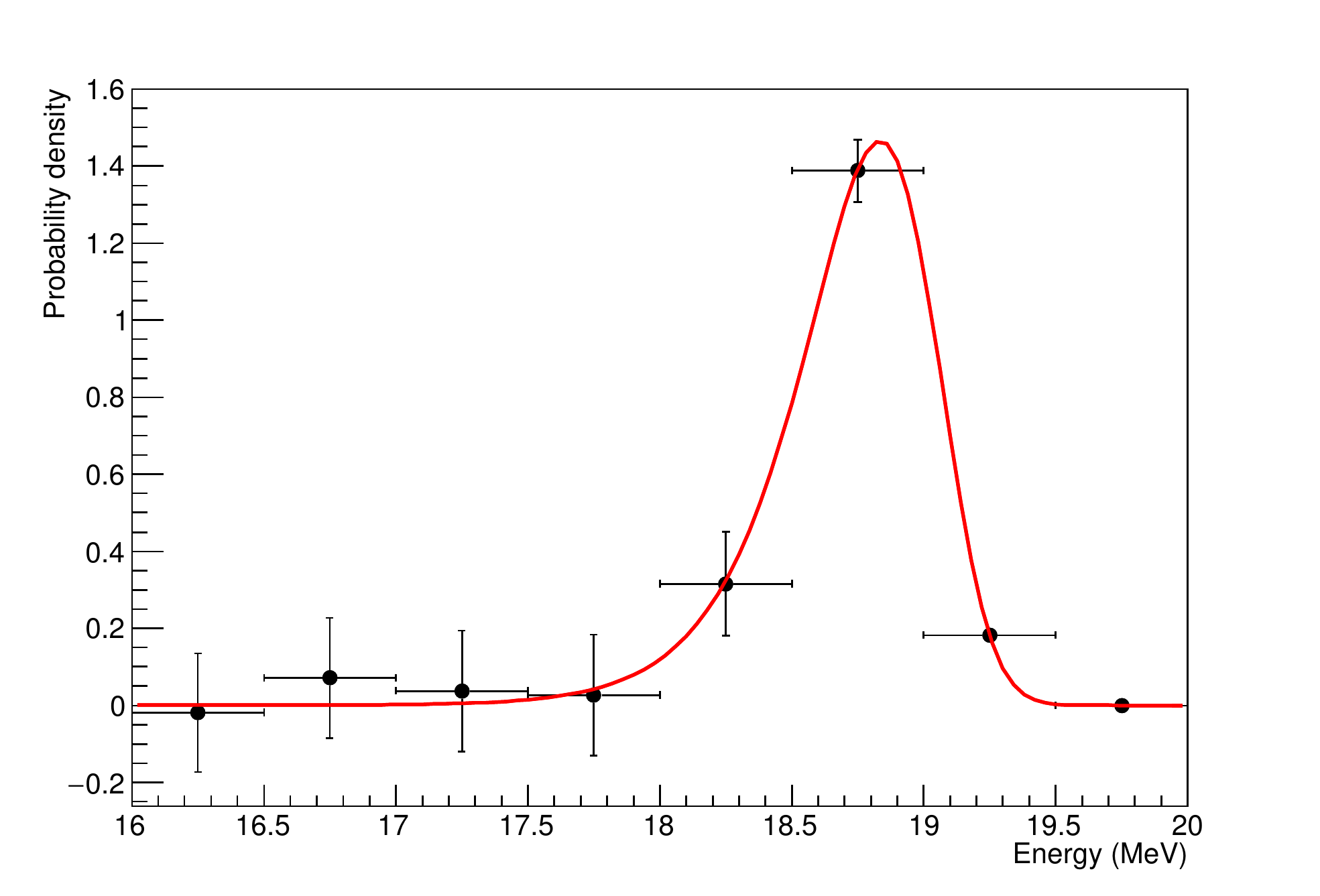}
   \caption{Distribution of the proton energy. The red line corresponds to the best fit of the Verhulst function.}
   \label{distribution}
\end{figure}

The results obtained for 9 aluminum absorbers are given in figure~\ref{distribution}. The probability density is shown for each bin of 0.5~MeV width. A mean energy of $(18.3\pm 0.3)$~MeV and an RMS of $(0.4\pm0.2)$~MeV were obtained. Additionally, a fit using a Verhulst function~\cite{verhulst} was performed with $\tilde{\chi}^2=0.6$. The fitted function was chosen due to the skewness of the measured distribution and it is defined as:
\begin{equation}
P(x)=\frac{1}{A\cdot B}\cdot \frac{\left(2^{A}-1\right)\cdot \exp\left(\frac{x-C}{B}\right)}{\left(1+\left(2^{A}-1\right)\cdot \exp\left(\frac{x-C}{B}\right)\right)^{\frac{A+1}{A}}},
\end{equation}
where $A, B$, and $C$ are the fit parameters. For the best fit they were found to be: $A=0.13\pm0.03$, $B=0.24\pm0.02$~MeV, and $C=18.76\pm0.02$~MeV. As expected, the mean beam energy at the BTL was found to be larger than the nominal one. This is due to the fact that a specific stripper holder is used which locates the stripper foil at a radius 5~mm larger with respect to the other outports~\cite{swan}.  

\subsection{Simulation of the Beam Transport Line}
\begin{figure}[!htb]
   \centering
   \includegraphics[width=0.497\textwidth]{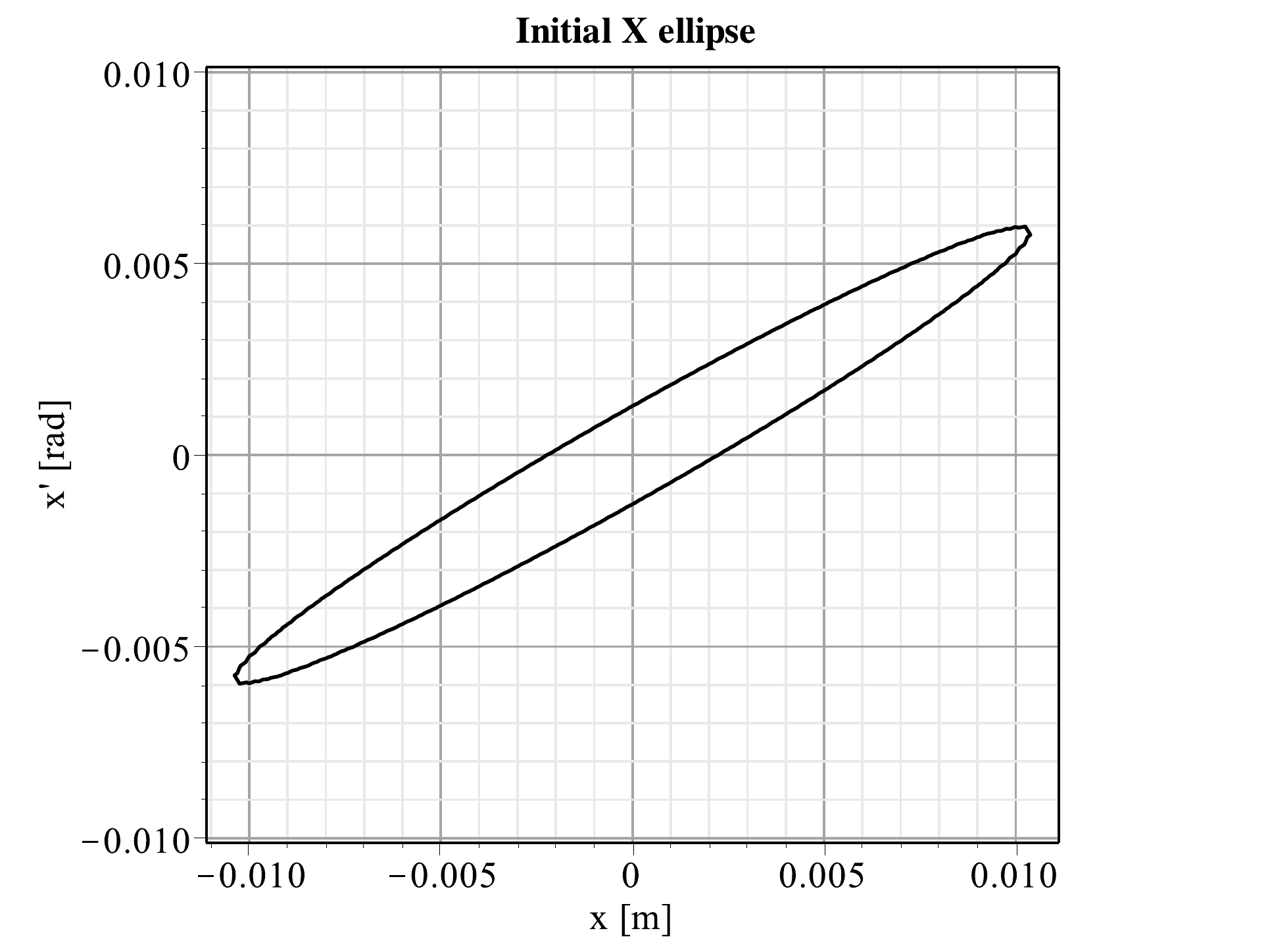}
      \includegraphics[width=0.497\textwidth]{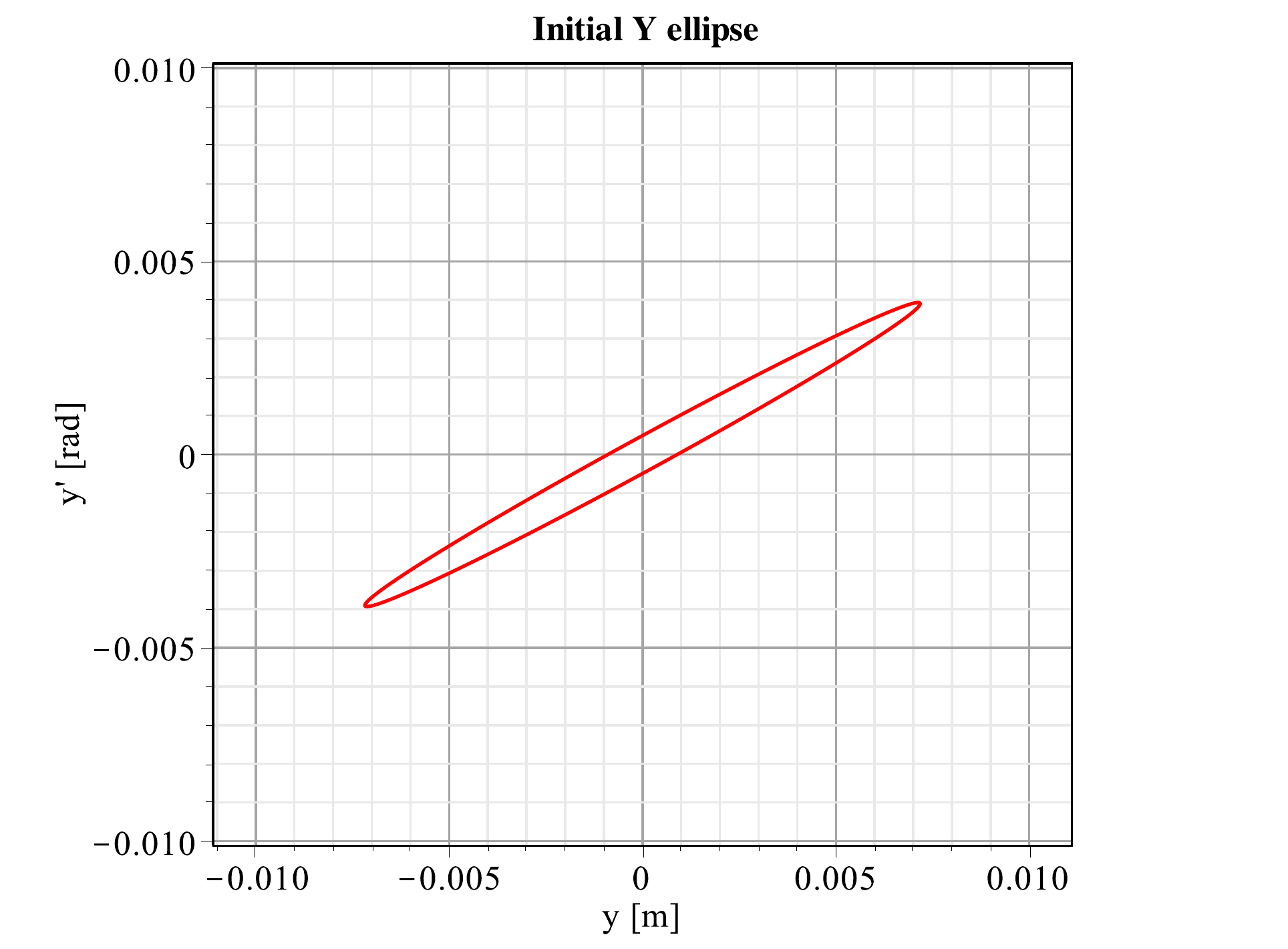}
   \caption{The horizontal (left) and vertical (right) 1$\sigma$-phase-space ellipses at the injection to the BTL.}
   \label{maple_ellipse}
\end{figure}

\begin{figure}[!htb]
   \centering
   \includegraphics[width=0.6\textwidth]{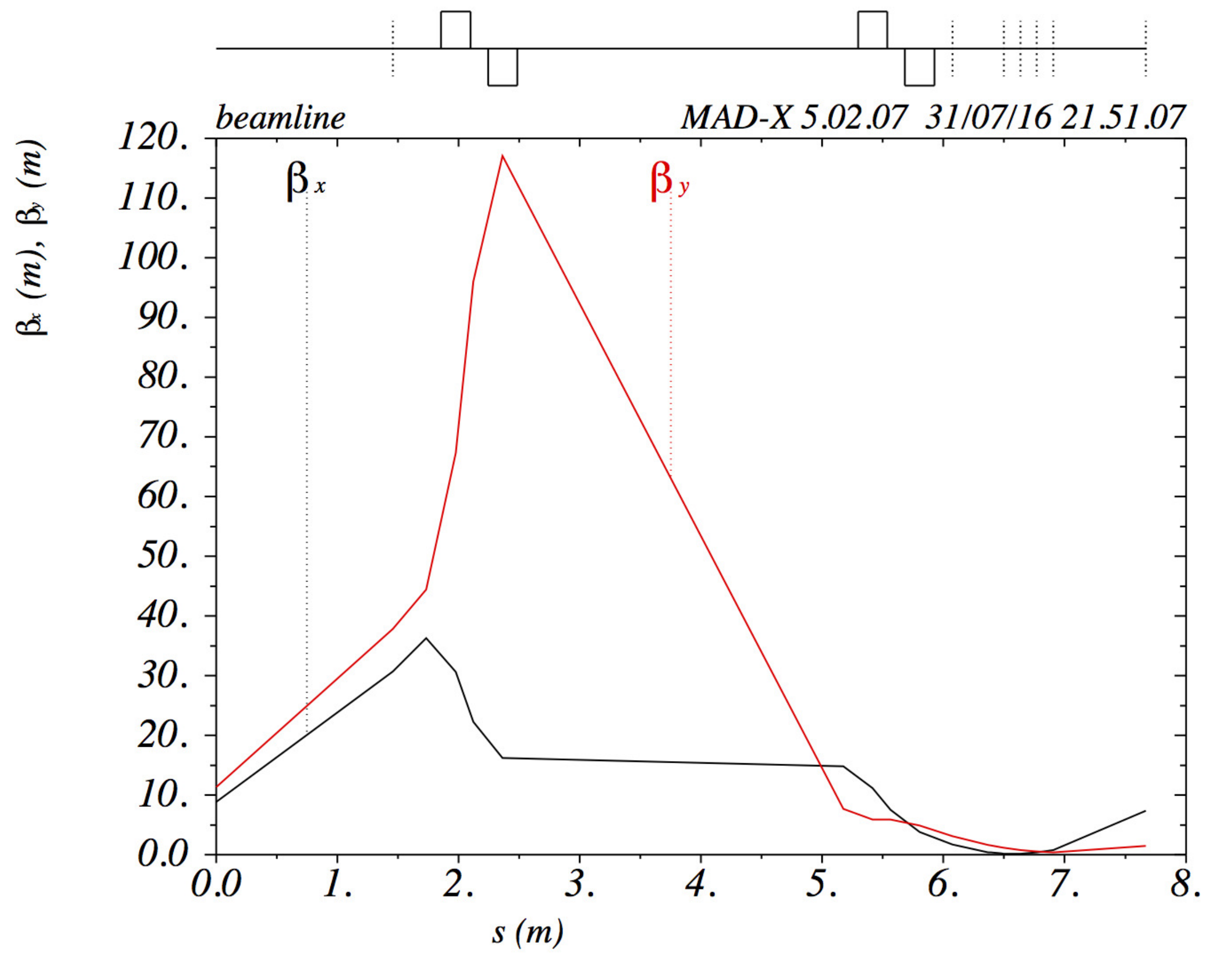}
   \caption{Simulated beam envelope in both horizontal and vertical planes.}
   \label{envelope}
\end{figure}

Methodical Accelerator Design (MAD-X) is a multi-purpose tool for charged-particle optics design and studies in alternating-gradient accelerators and beamlines. It was developed and is maintained by the Beams Department at CERN~\cite{madx}. It allows defining a beamline as a sequence of beam optics components, calculating Twiss parameters at their locations, and finding beamline component settings corresponding to specific constraints. A simulation of the BTL of the Bern cyclotron was implemented on the basis of the measurements performed with the multiple beam profiler method reported in this paper. Since the relative momentum spread was evaluated to be about 2 \%, it was not included in the simulation, having a negligible influence on beam envelopes. The Twiss parameters were calculated at the location of the first beam profiler and transported back to the beginning of the BTL by means of linear beam transport algebra. In this way the beam phase space at the injection to the BTL was reconstructed assuming Gaussian distributions of ($x$, $x'$) and ($y$, $y'$). The latter assumption leads to an ellipse limiting a certain fraction of the beam in the phase space.  As an example, the 1$\sigma$-ellipses for the standard cyclotron settings are shown in figure~\ref{maple_ellipse}. The simulation of the BTL is used for beam optimization in various experiments. In figure~\ref{envelope} a simulated beam envelope in both horizontal and vertical planes for the standard cyclotron parameters and standard quadrupole settings is reported. This tool is crucial to design experiments for radiation hardness studies, irradiation of solid targets, and radiobiological research activities.  

\section{Conclusions and outlook}
A system for online measurement of the transverse beam emittance, named $^{4}$PrOB$\varepsilon$aM, was developed at AEC-LHEP. It allows the transverse beam emittance to be measured in less than one minute. This compact system can be installed at any location along beamlines or directly at an accelerator outport. The characterization of the proton beam produced by the 18 MeV cyclotron in operation in Bern was performed. The transverse RMS beam emittance was measured with the $^{4}$PrOB$\varepsilon$aM system as a function of several cyclotron parameters. Such a scan of some crucial machine parameters was performed for the first time with a medical cyclotron and was possible due to a very short time of a single emittance measurement. Such measurements are essential for the beam optimization for multi-disciplinary research and can be useful for beamline commissioning and studies of machine imperfections. The transverse RMS emittance of the Bern cyclotron was also measured for the standard machine settings with two different methods and the results were found to be in a good agreement. The proton energy distribution at the BTL of the Bern cyclotron was also assessed for the first time. On the basis of the measurements reported in this paper, a simulation of the BTL was developed in the MAD-X software. This tool is essential for the optimization of the beams employed in the ongoing research activities.

The $^{4}$PrOB$\varepsilon$aM system can be deployed for similar measurements at other accelerator facilities. Further optimizations of the system, including simultaneous beam scanning in both planes are ongoing, and its commercialization is planned.

\acknowledgments
We acknowledge contributions from LHEP engineering and technical staff. The UniBEaM detector was partially developed in the framework of the grant by the Swiss National Science Foundation CR23I2\_156852.

\end{document}